\documentstyle[preprint,aps,graphicx]{revtex}
\newcommand{\be}{\begin{equation}}
\newcommand{\ee}{\end{equation}}
\newcommand{\bea}{\begin{eqnarray}}
\newcommand{\eea}{\end{eqnarray}}
\newcommand{\bref}[1]{(\ref{#1})}

\begin{document}
\draft
\title{Renormalization Group Equation of
 Quark-Lepton Mass Matrices in the SO(10) 
Model with Two Higgs Scalars}

\author{\bf Takeshi FUKUYAMA and Tatsuru KIKUCHI}
%\footnote{E-mail: koide@u-shizuoka-ken.ac.jp} \\
\address{
Department of Physics, Ritsumeikan University, 
Kusatsu, Shiga, 525-8577 Japan }
\date{\today}
\maketitle
\begin{abstract}
The renormalization group equations (RGEs) of the 
mass matrices of quarks and leptons in a SO(10) model
 with two Higgs scalars in the Yukawa coupling are studied. 
This model is the
 minimal model of SUSY and non-SUSY SO(10) GUT which can reproduce
 all the experimental data. Non-SUSY SO(10) GUT model has the
 intermediate energy phase, Pati-Salam phase, and passes through the symmetry
breaking pattern,
 $SO(10) \rightarrow SU(2)_L \times SU(2)_R \times SU(4)_C \rightarrow 
SU(2)_L \times U(1)_Y \times SU(3)_C$. 
Though minimal, it has, after the Pati-Salam phase, 
four Higgs doublets in Yukawa interactions.
%the minimal, neverthless 
%it is not so simple : each up-type quark-lepton and down-type
% quark-lepton couples to two Higgs doublets after the GUT breaking. 
%Enough below the intermediate scale,
% the effective neutrino mass operators are dominated by
%the dimention five operators. 
We consider the RGE's of the Yukawa coupling constants of 
quarks and charged leptons
 and of the coupling constants of the dimension five operators 
of neutrinos corresponding to the above symmetry breaking pattern. 
The scalar quartic interactions are also incorporated.
\end{abstract}

%%%%%%%%%%%%%%%%%%%%%%%%%%%%%%%%%%%%%%%%%%%
%\begin{multicols}{2}
In these decades many informations on the quark lepton mass
 matrices have been accumulated. We are confronted with the
 very era when we should seriously consider a realistic model
 in the scheme of grand unified theories.
 On this ocasion, neutrino masses may be the window to the
 grand unified theories via heavy right-handed neutrinos.
 Along this sight we considered in \cite{matsuda1} and 
\cite{matsuda2} the SO(10) model where two
 Higgs scalars participate in the Yukawa coupling. Our 
SO(10) model with two Higgs scalar, ${\bf 10}$ and 
$\overline{{\bf 126}}$,  may be the first realistic model 
which successfully fit with the quark lepton mass spectra,
 Cabbibo-Kobayashi-Maskawa (CKM) \cite{kobayashi}, and 
Maki-Nakagawa-Sakata-Pontecorvo (MNSP) \cite{maki} mixing matrices 
based on the SO(10) framework.  In these studies SO(10) 
invariant model is, of course, valid at the GUT scale and
 the data in our hand are those at much lower energy scale,
 at electroweak scale.  So we must transport either of them
 to the other energy scale using the renormalization group
 equations (RGEs). In \cite{fukuyama} we bottomed up the 
electroweak scale data about six quark masses, three mixing
 angles and one CP-phase in the CKM matrix and three charged
 lepton masses to the GUT scale and, using the SO(10) GUT
 mass relations (see Eqs.\bref{triplet} and (\ref{mass2}))
, obtained the neutrino mass matrix via seesaw mechanics 
\cite{yanagida} at GUT or at the intermediate scale. Then we toped
 down this mass matrix to the electroweak scale
 and checked whether it is consistent with the neutrino
 oscillation data and neutrinoless double beta decay. 
The results were very satisfactory in principle.
 However the theory is, of course, not conclusive.  We adopted 
there some assumtions in order to extract the essential characters.
 One of them was the doublet-doblet splitting:
 that is, one of doublets is heavy relative to the other and our
 model in \cite{fukuyama} was the same in essence as the MSSM
 except for the additional mixing angle among heavy and light Higgs doublets.
However, this doublet-doublet splitting is not sure to occur.
 For instance, in the same SO(10) GUT with two Higgs scalar, these 
two Higgs doublet may work in parallel to remedy the over-abundance of
 the leptogenesis \cite{okada}.
So in this paper we discarded this assumption and also incorporate 
scalar self coupling. In pay of this generalization we abandon
 in this paper the detailed data fitting and restrict
 ourselves in showing RGE's of quark-leptons mass 
matrices on more general ground. 
The renormalization of the neutrino mass operator 
was discussed by Babu-Leung-Pantaleone
 and Chankowski-Pluciennik \cite{babu} for some simple models.
 Unfortunately it includes some calculational errors \cite{antusch},
 though it does not affect our work, because we assume 
nuetrino couples to the Higgs doublets in a similar way
 to the up quarks. Our model is the most simple realistic
 model in SUSY and non-SUSY SO(10) GUT. 
At the GUT scale we have the Yukawa interactions to be given by
\begin{eqnarray}
{\mathcal L}_{Y} =\sum_{i,j}\left( 
Y^{(10)}_{ij} \Psi^{iT}B \gamma^{\mu} {H_{\mu}} \Psi^{j} 
+\frac{1}{\left(5!\right)^2}Y^{(126)}_{ij} \Psi^{iT}B 
\gamma^{\mu_1}\gamma^{\mu_2}\gamma^{\mu_3}\gamma^{\mu_4}\gamma^{\mu_5} 
{\overline{\Delta}}_{\mu_1 \mu_2 \mu_3 \mu_4 \mu_5}
 \Psi^{j}+h.c. \right), 
\label{Yukawa1}
\end{eqnarray} 
where $\Psi^{i}\left(i=1,2,3 \right)$ is the
 {\bf 16}-dimentional matter multiplet  of the 
i-th generation, $H$ and $\overline{\Delta}$ are
 the Higgs multiplet  of {\bf 10} and $\overline{\bf 126}$
 representations under SO(10), respectively. $B$
 denotes the charge conjugation for SO(10) spinors
 : $B=\gamma_{1}\gamma_{3}\gamma_{5}\gamma_{7}\gamma_{9}$. \\
The Higgs scalar's quartic interactions with only
 $H$ and $\overline{\Delta}$ are given by
\begin{eqnarray}
 V\left(H,\overline{\Delta} \right)
&=&
 \lambda_1\left(H^{\mu}H_{\mu} \right)\left(H^{\nu}H_{\nu} \right)
+\frac{1}{5!}\lambda_2
\left(H^{\mu}H_{\mu} \right)
\left(\overline{\Delta}^{\mu_1 \mu_2 \mu_3 \mu_4 \mu_5}
\overline{\Delta}_{\mu_1 \mu_2 \mu_3 \mu_4 \mu_5}  \right)
 \nonumber\\
&+&\frac{1}{4!}\lambda_3\left(H^{\mu_1}
\overline{\Delta}_{\mu_1 \mu_2 \mu_3 \mu_4 \mu_5} \right)
\left(H_{\nu_1 }\overline{\Delta}^{\nu_1 \mu_2 \mu_3 \mu_4 \mu_5}
 \right) \nonumber\\
&+&\frac{1}{\left(5!\right)^2}\lambda_4
\left(\overline{\Delta}^{\mu_1 \mu_2 \mu_3 \mu_4 \mu_5}
\overline{\Delta}_{\mu_1 \mu_2 \mu_3 \mu_4 \mu_5} \right)
\left(\overline{\Delta}^{\nu_1 \nu_2 \nu_3 \nu_4 \nu_5}
\overline{\Delta}_{\nu_1 \nu_2 \nu_3 \nu_4 \nu_5} \right)+h.c.
 \label{guthiggspotential}
\end{eqnarray}
Hereafter we consider non-SUSY SO(10) GUT explicitly. Gauge coupling 
unification needs the intermediate energy scale, $\Lambda_I$ \cite{pati}.
Between the grand unification scale and the intermediate
 scale, the effective Yukawa interactions are given by
\begin{eqnarray}
 -{\mathcal L}_{Y} &=&\sum_{i,j}\left( Y^{(10)}_{F\,ij}F_L^{iT}\Phi F_R^{j}
  +Y^{(126)}_{F\,ij}F_L^{iT}\Sigma F_R^{j}
+Y^{(126)}_{R\,ij}F_R^{iT}\overline{\Delta_R} F_R^{j}+h.c.\right),
\label{y2}
\end{eqnarray}
where $F_L$ and $F_R$ denote $\bf{(2,1,4)}$ and 
$\bf{(1,2,{\overline{4}})}$ in $\Psi^{i}$, under 
$G_{224} \equiv SU(2)_L \times SU(2)_R \times SU(4)_C$,
 respectively. And also, $\Phi$, $\Sigma$ and $\overline{\Delta_R}$ 
correspond 
to $\bf{(2, 2, 1)}$ in $H$, $\bf{(2, 2, 15)}$ and $\bf{(1, 3, \overline{10})}$ 
in $\overline{\Delta}$, respectively.  
Here we have assumed that suitably chosen $U(1)_H$ charge forbids the 
Yukawa interactions like, $F_L^{i\,T} \widetilde{\Phi} F_R^{j}$ and 
$F_L^{i\,T} \widetilde{\Sigma} F_R^{j}$,  where \\
\begin{equation}
\widetilde{\Phi}=\epsilon^{T} \Phi^{*} \epsilon, \quad 
\widetilde{\Sigma}=\epsilon^{T} \Sigma^{*} \epsilon, 
\end{equation}
with
\begin{equation}
\epsilon=
\left(
\begin{array}{ll}
\ 0 & \ 1 \\
\ -1 & \ 0 \\
\end{array}
\right).
\end{equation}
And also the Higgs scalar's quartic interactions are given by 
\begin{eqnarray}
V\left(\Phi,\Sigma \right)
&=&
 \lambda_1\left[{\mathrm{tr}}\left(\Phi \Phi^{\dagger}\right)\right]^2
+2 \lambda_2\,{\mathrm{tr}}\left(\Phi \Phi^{\dagger}\right)
{\mathrm{tr}}\left(\Sigma \Sigma^{\dagger}\right)
\nonumber\\
&+&\lambda_3\left\{
\left[{\mathrm{tr}}\left(\Phi \Sigma^{\dagger}\right)\right]^2
+\left[{\mathrm{tr}}\left(\Sigma \Phi^{\dagger}\right)\right]^2\right\}
+4 \lambda_4\left[{\mathrm{tr}}\left(\Sigma \Sigma^{\dagger}\right)\right]^2.
\label{interhiggspotential}
\end{eqnarray}
When the intermediate symmetry 
breaking occures, $\overline{\Delta_R}$ have the vacuum 
expectation value and 
\begin{equation}
M_R=\left\langle \overline{\Delta_R} 
\right\rangle Y^{(126)}_R.
\label{triplet}
\end{equation}
Below the intermediate scale, it includes 
four Higgs doublets : two doublets $(\phi_1$ and $\phi_3)$ 
come from $\bf{(2, 2, 1)}$ and $(\phi_2$ and $\phi_4)$  from 
$\bf{(2, 2, 15)}$. $\{\phi_1$, $\phi_2\}$ couple with up-type 
quarks and leptons, and  $\{\phi_3$, $\phi_4\}$ do with down-type 
quarks and leptons \cite{matsuda3}.
  In these multiple Higgs models we always encounter with
the flavor changing neutral current (FCNC) problem as Glashaw 
and Weinberg remarked \cite{glashow}.
However, by choosing the specific mass matrix textures, 
FCNC can be suppressed even in the case of multiple 
Higgs case \cite{sher}. 
Thus the low energy
 effective Yukawa interactions and the Higgs scalar's quartic
 interactions have the following forms,
\begin{eqnarray}
-{\mathcal L}_{Y}&=&\sum_{i,j}\left(Y^{(10)}_{u\,ij}\,
\overline{q^i_L}\,\widetilde{\phi_1}u_R^j
+Y^{(126)}_{u\,ij}\,\overline{q^i_L}\,\widetilde{\phi_2}u_R^j 
\right.\nonumber\\
&+&Y^{(10)}_{d\,ij}\,\overline{q^i_L}\,\phi_3d_R^j
+Y^{(126)}_{d\,ij}\,\overline{q^i_L}\,\phi_4d_R^j \nonumber\\
&+&Y^{(10)}_{\nu\, ij}\,\overline{\ell^i_L}\,\widetilde{\phi_1}N_R^j
+Y^{(126)}_{\nu\, ij}\,\overline{\ell^i_L}\,\widetilde{\phi_2}N_R^j
 \nonumber\\
&+&\left.Y^{(10)}_{e\,ij}\,\overline{\ell^i_L}\,\phi_3e_R^j
+Y^{(126)}_{e\,ij}\,\overline{\ell^i_L}\,\phi_4e_R^j+h.c.
\right) \nonumber\\
&+&\frac{1}{2}\sum_{i,j}\left(M_{R\,ij}\overline{N^{ci}_R}N^j_R
+h.c. \right),
\label{y}
\end{eqnarray}
\begin{eqnarray}
 V\left(\phi \right)&=&\frac{1}{4!}\sum_{a,b.c.d=1,2,3,4}
\lambda_{abcd} 
\left(\phi_a^{\dagger} \phi_b \right)
\left(\phi_c^{\dagger} \phi_d \right).
\label{higgspotential}
\end{eqnarray}
 Here $q_L$ and $\ell_L$ are the left-handed quark and lepton doublets : 
\begin{equation}
q_{L} =\left(
\begin{array}{c}
u_L \\
d_L
\end{array} \right) \ , \ \ \ 
\ell_{L} =\left(
\begin{array}{c}
\nu_{L} \\
e_{L}
\end{array} \right) \ .
\end{equation}
Also, $\phi_{a}$ are the Higgs doublets :
\begin{equation}
\phi_{a} =\left(
\begin{array}{c}
\phi^+_{a} \\
\phi^0_{a}
\end{array} \right) \ , \ \ \ 
\widetilde{\phi}_{a} =\left(
\begin{array}{c}
\phi^{0*}_{a} \\
-\phi^-_{a}
\end{array} \right) \ . \ \ \
\end{equation}
From the hermiticity of \bref{higgspotential}, the self coupling constants $\lambda_{abcd}$ satisfy
\be
\lambda_{abcd}=\lambda_{cdab}=\lambda_{badc}^*.
\ee
Eq.(\ref{y2}) forbids the Yukawa and the scalar quartic interactions 
in Eqs.(\ref{y}) and (\ref{higgspotential}) which are not invariant under
the $\mathbf{Z}_2$ symmetry,
\begin{eqnarray}
 \phi_{1}\to \phi_{1}, \,\,\, \phi_{2}\to \phi_{2}, 
\nonumber\\
 \phi_{3}\to -\phi_{3}, \,\,\, \phi_{4}\to -\phi_{4}
\end{eqnarray}
and corresponding transformations in the fermion sector. 
When the electroweak symmetry breaking occures,
 $\phi^{0}_a~(a=1,2,3,4)$ have the vacuum expectation values(VEVs),
\begin{equation}
\left\langle \phi^{0}_a \right\rangle 
=\frac{v_a}{\sqrt{2}}~~~(a=1,2,3,4)
\end{equation}
and 
\begin{eqnarray}
M_u&=&\frac{v_1^{*}}{\sqrt{2}}Y^{(10)}_u
+\frac{v_2^{*}}{\sqrt{2}}Y^{(126)}_u,\nonumber\\
M_d&=&\frac{v_3}{\sqrt{2}}Y^{(10)}_d
+\frac{v_4}{\sqrt{2}}Y^{(126)}_d,\nonumber\\
M_D&=&\frac{v_1^{*}}{\sqrt{2}}Y^{(10)}_{\nu}
+\frac{v_2^{*}}{\sqrt{2}}Y^{(126)}_{\nu},\nonumber\\M_e&=&\frac{v_3}
{\sqrt{2}}Y^{(10)}_e+\frac{v_4}{\sqrt{2}}Y^{(126)}_e.
\label{mass}
\end{eqnarray}
The Yukawa couplings above the intermediate energy scale are unified to
\begin{eqnarray}
\lefteqn{\frac{1}{\sqrt{2}}Y^{(10)}_F \equiv Y^{(10)},} \nonumber\\
&&\frac{1}{4\sqrt{2}}Y^{(126)}_F=\frac{1}{4}Y^{(126)}_R 
\equiv Y^{(126)}
\label{bc1}
\end{eqnarray}
at the GUT scale. 
Here the above numerical factors in Eq.(\ref{bc1}) are necessarily 
to translate 
the $SO(10)$ language into the $G_{224}$ language, see for example 
\cite{aulakh}. 
And also, at the the intermediate scale, 
the Yukawa couplings above the electroweak scale are unified to
\begin{eqnarray}
\lefteqn{Y^{(10)}_u=Y^{(10)}_d=Y^{(10)}_{\nu}
=Y^{(10)}_e\equiv Y^{(10)}_F,} \nonumber\\
&&Y^{(126)}_u=Y^{(126)}_d=-\frac{1}{3}Y^{(126)}_{\nu}
=-\frac{1}{3}Y^{(126)}_e\equiv Y^{(126)}_F,
\end{eqnarray}
and Eq.(\ref{mass}) is reduced to \cite{rabi}
\begin{eqnarray}
M_u&=&c_{10} M^{(10)}+c_{126} M^{(126)},
~~M_d=M^{(10)}+M^{(126)},\nonumber\\
M_{D}&=&c_{10} M^{(10)}-3c_{126} M^{(126)},
~~M_e=M^{(10)}-3M^{(126)}.
\label{mass2}
\end{eqnarray}
Here 
\be
c_{10} \equiv v_1^{*}/v_3\,,~~~~c_{126} \equiv v_2^{*}/v_4\,,
\ee
and
\begin{equation}
M^{(10)} \equiv \frac{v_3}{\sqrt{2}}\,Y^{(10)}_F,
~~~M^{(126)} \equiv \frac{v_4}{\sqrt{2}}\,Y^{(126)}_F.
\end{equation}
It goes from Eq.(\ref{interhiggspotential}) that the Higgs 
quartic coupling constants are unified to
\bea
\frac{\lambda_{abcd}}{4!}&=&
\lambda_1~~\mbox{for}~a,b,c,d=1,3~\mbox{only}.\nonumber\\
\frac{\lambda_{abcd}}{4!}&=&2
\lambda_2~~\mbox{for}~a,b=1,3~\mbox{and}~c,d=2,4.\\
\frac{\lambda_{abcd}}{4!}&=&
\lambda_3~~\mbox{for}~\{a,b\}~\mbox{and}~\{c,d\}\,\,
 \mbox{take any sets of}~\{1,2\},~\{1,4\},~\{3,2\},~\{3,4\}.\nonumber\\
\frac{\lambda_{abcd}}{4!}&=&4
\lambda_4~~\mbox{for}~a,b,c,d=2,4~\mbox{only}\nonumber
\eea
at the intermediate scale.  Here the coupling constants which does 
not satisfy Eqs.(\ref{guthiggspotential}) are vanished 
and excluded from the above rule.
It should be emphasized that this model can be 
compatible with the large angle of atmospheric 
neutrino oscillation as far as we do not adopt 
any simplification not allowed in SO(10) framework 
\cite{matsuda2}. For $\Lambda_{I} \gg \mu$, $M_R$ decouples 
and we must treat the four point interaction,
\begin{eqnarray}
{\mathcal L}_{\nu \nu}&=&\frac{1}{4}\sum_{a,b=1,2}
\sum_{i,j}\kappa^{(a,b)}_{ij}\left(\overline{\ell^i_L}
\widetilde{\phi_{a}}\right)\left(\phi^{*}_{b\alpha}
\epsilon^{\alpha\beta}\ell^{Cj}_{L\beta}\right)+h.c. 
\nonumber\\
&=&-\frac{1}{2}\sum_{a,b=1,2}\sum_{i,j}\kappa^{(a,b)}_{ij}
\left\{\left(\overline{\nu^i_L}\nu^{Cj}_L\phi^{0*}_a\phi^{0*}_b
+\overline{e^i_L}e^{Cj}_L\phi^-_a\phi^-_b\right) \right.
 \nonumber\\
&-&\left.\frac{1}{2}\left(\overline{\nu^i_L}e^{Cj}_L
+\overline{e^i_L}\nu^{Cj}_L \right)
\left(\phi^-_a\phi^{0*}_b+\phi^{0*}_a\phi^-_b\right)\right\}+h.c.,
\label{fourpoint}
\end{eqnarray}
where $i,j$ are flavour indicies. The reason why $a,b$ 
run over $1,2$ is that it comes from the Dirac neutrinos 
(the third line in the right-hand side of Eq.(5)). 
Thus the effective light neutrino mass matrix is given by
\begin{equation}
M_{\nu}=\frac{1}{2}\sum_{a,b=1,2}\kappa^{(a,b)}v_a^{*}v_b^{*}.
\label{Mnu}
\end{equation}
We have two Higgs doublets in each up quarks, down quarks,
 Dirac neutrinos and charged leptons and 
$\kappa^{(a,b)}~~(a,b=1,2)$ in ours are the 
generaslization of $\kappa^{(22)}$ in \cite{babu}.
 We have no term corresponding to $\kappa^{(11)}$ in \cite{babu}.\\
In the non SUSY SO(10) with two Higgs scalars, the one 
loop RGEs for the effective Yukawa couplings first at 
the energy region between the grand 
unification scale and the intermediate sale are given by :
\begin{eqnarray}
16\pi^2\frac{dY^{(10)}_F}{dt}&=&\left(Y^{(10)}_FY^{(10)\dagger}_F
+\frac{15}{4}Y^{(126)}_FY^{(126)\dagger}_F\right)Y^{(10)}_F \nonumber\\
&+&Y^{(10)}_F\left\{Y^{(10)}_FY^{(10)\dagger}_F
+\frac{15}{4}\left(Y^{(126)}_F Y^{(126)\dagger}_F
+Y^{(126)}_R Y^{(126)\dagger}_R \right)\right\} \nonumber\\
&+&4{\mathrm{tr}}\left(Y^{(10)}_F Y^{(10)\dagger}_F\right)Y^{(10)}_F
+\left(\frac{9}{4}g_{2L}^{2}+\frac{9}{4}g_{2R}^{2}
+\frac{15}{4}g_{4C}^{\,2}\right) Y^{(10)}_F,\\
16\pi^2\frac{dY^{(126)}_F}{dt}&=&\left(Y^{(10)}_FY^{(10)\dagger}_F
+\frac{15}{4}Y^{(126)}_FY^{(126)\dagger}_F\right)Y^{(126)}_F \nonumber\\
&+&Y^{(126)}_F\left\{Y^{(10)}_FY^{(10)\dagger}_F
+\frac{15}{4}\left(Y^{(126)}_FY^{(126)\dagger}_F
+Y^{(126)}_R Y^{(126)\dagger}_R \right)\right\} \nonumber\\
&+&{\mathrm{tr}}\left(Y^{(126)}_FY^{(126)\dagger}_F\right)Y^{(126)}_F
+\left(\frac{9}{4}g_{2L}^{2}+\frac{9}{4}g_{2R}^{2}
+\frac{15}{4}g_{4C}^{2}\right) Y^{(126)}_F,\\
16\pi^2\frac{dY^{(126)}_R}{dt}&=&\left\{Y^{(10)}_FY^{(10)\dagger}_F
+\frac{15}{4}\left(Y^{(126)}_FY^{(126)\dagger}_F
+Y^{(126)}_R Y^{(126)\dagger}_R \right)\right\}Y^{(126)}_R \nonumber\\
&+&Y^{(126)}_R\left\{Y^{(10)}_FY^{(10)\dagger}_F
+\frac{15}{4}\left(Y^{(126)}_FY^{(126)\dagger}_F
+Y^{(126)}_R Y^{(126)\dagger}_R \right)\right\} \nonumber\\
&+&{\mathrm{tr}}\left(Y^{(126)}_RY^{(126)\dagger}_R\right)Y^{(126)}_R
+\left(\frac{9}{2}g_{2R}^{2}
+\frac{15}{4}g_{4C}^{2}\right) Y^{(126)}_R,
\end{eqnarray}
where $g_{2L}$, $g_{2R}$ and $g_{4C}$ are the $SU(2)_L$
, $SU(2)_R$ and $SU(4)_C$ 
gauge coupling constants, respectively. 
At the second stage, the energy region 
between the intermediate sale and the weak scale, 
the one loop RGEs for the effective Yukawa couplings are given by :
\begin{eqnarray}
16\pi^2\frac{dY^{(10)}_u}{dt}&=&3{\mathrm{tr}}
(Y^{(10)}_uY^{(10)\dagger}_u)Y^{(10)}_u
+3{\mathrm{tr}}(Y^{(10)}_uY^{(126)\dagger}_u)Y^{(126)}_u \nonumber\\
&-&\left(8g_3^2+\frac{9}{4}g_2^2
+\frac{17}{12}g_Y^2\right)Y_u^{(10)} \nonumber\\
&+&{1 \over 2}\left(Y^{(10)}_uY^{(10)\dagger}_u 
+Y^{(126)}_uY^{(126)\dagger}_u \right.\nonumber\\
&+&\left. Y^{(10)}_dY^{(10)\dagger}_d
+Y^{(126)}_dY^{(126)\dagger}_d \right)Y^{(10)}_u \nonumber\\
&+&Y^{(10)}_u\left(Y^{(10)\dagger}_uY^{(10)}_u
 +Y^{(126)\dagger}_uY^{(126)}_u\right),\label{yu10} \\
16\pi^2\frac{dY^{(126)}_u}{dt}&=&3{\mathrm{tr}}
(Y^{(126)}_uY^{(126)\dagger}_u)Y^{(126)}_u
+3{\mathrm{tr}}(Y^{(126)}_uY^{(10)\dagger}_u)Y^{(10)}_u \nonumber\\
&-&\left(8g_3^2+\frac{9}{4}g_2^2+\frac{17}{12}g_Y^2\right)
Y_u^{(126)} \nonumber\\
&+&{1 \over 2}\left(Y^{(10)}_uY^{(10)\dagger}_u
+Y^{(126)}_uY^{(126)\dagger}_u \right. \nonumber\\
&+&\left. Y^{(10)}_dY^{(10)\dagger}_d
+Y^{(126)}_dY^{(126)\dagger}_d \right)Y^{(126)}_u \nonumber\\
&+&Y^{(126)}_u\left(Y^{(10)\dagger}_uY^{(10)}_u 
+Y^{(126)\dagger}_uY^{(126)}_u\right),\\
16\pi^2\frac{dY^{(10)}_d}{dt}&=&\left\{3{\mathrm{tr}}
(Y^{(10)}_dY^{(10)\dagger}_d)+{\mathrm{tr}}
(Y^{(10)}_eY^{(10)\dagger}_e)\right\}Y^{(10)}_d \nonumber\\
&+&\left\{3{\mathrm{tr}}(Y^{(10)}_dY^{(126)\dagger}_d)
+{\mathrm{tr}}(Y^{(10)}_eY^{(126)\dagger}_e)\right\}Y^{(126)}_d \nonumber\\
&-&\left(8g_3^2+\frac{9}{4}g_2^2
+\frac{5}{12}g_Y^2\right)Y_d^{(10)} \nonumber\\
&+&{1 \over 2}\left(Y^{(10)}_uY^{(10)\dagger}_u 
+ Y^{(126)}_uY^{(126)\dagger}_u \right. \nonumber\\
&+&\left. Y^{(10)}_dY^{(10)\dagger}_d 
+Y^{(126)}_dY^{(126)\dagger}_d \right)Y^{(10)}_d  \nonumber\\
&+&Y^{(10)}_d\left(Y^{(10)\dagger}_dY^{(10)}_d 
+Y^{(126)\dagger}_dY^{(126)}_d\right),\\
16\pi^2\frac{dY^{(126)}_d}{dt}&=&\left\{3{\mathrm{tr}}
(Y^{(126)}_dY^{(126)\dagger}_d)+{\mathrm{tr}}
(Y^{(126)}_eY^{(126)\dagger}_e)\right\}Y^{(126)}_d \nonumber\\
&+&\left\{3{\mathrm{tr}}(Y^{(126)}_dY^{(10)\dagger}_d)
+{\mathrm{tr}}(Y^{(126)}_eY^{(10)\dagger}_e)\right\}Y^{(10)}_d \nonumber\\
&-&\left(8g_3^2+\frac{9}{4}g_2^2
+\frac{5}{12}g_Y^2\right)Y_d^{(126)} \nonumber\\
&+&{1 \over 2}\left(Y^{(10)}_uY^{(10)\dagger}_u 
+Y^{(126)}_uY^{(126)\dagger}_u \right. \nonumber\\
&+&\left. Y^{(10)}_dY^{(10)\dagger}_d 
+Y^{(126)}_dY^{(126)\dagger}_d \right)Y^{(126)}_d  \nonumber\\
&+&Y^{(126)}_d\left(Y^{(10)\dagger}_dY^{(10)}_d 
+Y^{(126)\dagger}_dY^{(126)}_d\right),\\
16\pi^2\frac{dY^{(10)}_e}{dt}&=&\left\{3{\mathrm{tr}}
(Y^{(10)}_dY^{(10)\dagger}_d)+{\mathrm{tr}}
(Y^{(10)}_eY^{(10)\dagger}_e)\right\}Y^{(10)}_e \nonumber\\
&+&\left\{3{\mathrm{tr}}(Y^{(10)}_dY^{(126)\dagger}_d)
+{\mathrm{tr}}(Y^{(10)}_eY^{(126)\dagger}_e)\right\}Y^{(126)}_e \nonumber\\
&-&\left(\frac{9}{4}g_2^2+\frac{15}{4}g_Y^2\right)Y_e^{(10)} \nonumber\\
&+&{1 \over 2}\left(Y^{(10)}_eY^{(10)\dagger}_e 
+Y^{(126)}_eY^{(126)\dagger}_e \right)Y^{(10)}_e \nonumber\\
&+&Y^{(10)}_e\left(Y^{(10)\dagger}_eY^{(10)}_e
+Y^{(126)\dagger}_eY^{(126)}_e\right),\\
16\pi^2\frac{dY^{(126)}_e}{dt}&=&\left\{3{\mathrm{tr}}
(Y^{(126)}_dY^{(126)\dagger}_d)+{\mathrm{tr}}
(Y^{(126)}_eY^{(126)\dagger}_e)\right\}Y^{(126)}_e \nonumber\\
&+&\left\{3{\mathrm{tr}}(Y^{(126)}_dY^{(10)\dagger}_d)
+{\mathrm{tr}}(Y^{(126)}_eY^{(10)\dagger}_e)\right\}Y^{(10)}_e \nonumber\\ 
&-&\left(\frac{9}{4}g_2^2+\frac{15}{4}g_Y^2\right)
Y_e^{(126)} \nonumber\\
&+&{1 \over 2}\left(Y^{(10)}_eY^{(10)\dagger}_e
+Y^{(126)}_eY^{(126)\dagger}_e \right)Y^{(126)}_e \nonumber\\
&+&Y^{(126)}_e\left(Y^{(10)\dagger}_eY^{(10)}_e
+Y^{(126)\dagger}_eY^{(126)}_e\right).\label{ye126}
\end{eqnarray}
In Eqs.(\ref{yu10})-(\ref{ye126}), the first two terms 
(before $g_i^2$ terms) are the contribution of fermion 
loop, the third term that of gauge loop, and the 
remaining two terms those of Higgs loop. The second 
term indicates the mixing of two Higgs doublets in 
the fermion loop correction.\\
These formulas, of course, are reduced to one-loop 
RGEs for the standard model if we set either 
$Y^{(10)}$ or $Y^{(126)}$ zero \cite{arason}. 
The one loop RGEs for the scalar quartic couplings 
in our model are \cite{cheng}
\begin{eqnarray}
16\pi^2\frac{d\lambda_{abcd}}{dt}&=&\frac{1}{6}\sum_{m,n=1,2,3,4}
\left(2\lambda_{abmn} \lambda_{nmcd}+\lambda_{abmn} \lambda_{cmnd}
+\lambda_{amnb} \lambda_{mncd} \right. \nonumber\\
&+&\left. \lambda_{amnd} \lambda_{cnmb}+\lambda_{amcn} \lambda_{mbnd}\right)
-3\left(3g_2^2+g_Y^2 \right)\lambda_{abcd} \nonumber\\
&+&9\left(3g_2^4+g_Y^4\right)\delta_{ab}\delta_{cd}
+36 \, g_2^2 g_Y^2 \left(\delta_{ad}\delta_{bc}
-\frac{1}{2}\delta_{ab}\delta_{cd} \right) \nonumber\\
&+&\sum_{m,n=1,2,3,4}\left(\lambda_{mbcd} A_{am}+\lambda_{amcd} A_{mb}
+\lambda_{abmd} A_{cm}+\lambda_{abcm} A_{md}\right)
 \nonumber\\
&-&48 \, H_{abcd}
\quad \left(a,b,c,d=1,2,3,4 \right),
\label{scalarrge}
\end{eqnarray}
Here
\be
A_{ab} \equiv 
{\mathrm{tr}}\left(3Y_{a}^{u \dagger}Y_{b}^{u }+
3Y_{a}^{d \dagger}Y_{b}^{d }+
Y_{a}^{e \dagger}Y_{b}^{e }\right),
\ee
and
\bea
H_{abcd}&\equiv &
{\mathrm{tr}}\left(3Y_{d}^{u \dagger}Y_{c}^{u }Y_{b}^{u \dagger}Y_{a}^{u }
+3Y_{a}^{d \dagger}Y_{b}^{d }Y_{c}^{d \dagger}Y_{d}^{d }
+Y_{a}^{e \dagger}Y_{b}^{e }Y_{c}^{e \dagger}Y_{d}^{e }
\right. \nonumber\\
&+&\left. 
 3Y_{a}^{u \dagger}Y_{b}^{u }Y_{d}^{d \dagger}Y_{c}^{d }
+3Y_{b}^{d \dagger}Y_{a}^{d }Y_{c}^{u \dagger}Y_{d}^{u }
-3Y_{d}^{d \dagger}Y_{c}^{d }Y_{b}^{u \dagger}Y_{a}^{u }
-3Y_{a}^{u \dagger}Y_{d}^{u }Y_{b}^{d \dagger}Y_{c}^{d }
 \right),
\eea
with
\bea
Y_{1}^{u}&=&Y_u^{(10)},~~Y_{2}^{u}=Y_u^{(126)}, \quad
Y_{3}^{d}=Y_d^{(10)},~~Y_{4}^{d}=Y_d^{(126)},\nonumber\\
Y_{3}^{e}&=&Y_e^{(10)},~~Y_{4}^{e}=Y_e^{(126)},\quad
\mbox{and otherwise zero.}
\eea 
As for the RGEs of VEV's of Higgs fields, there 
may be some conflicts.
Someone consider them constants \cite{fukuyama} \cite{koide} \cite{ellis}, 
and other ones make them evolve as $\frac{d\sqrt{-2\mu^2/\lambda}}{dt}$ 
for the simplest case \cite{chankowski}. The situation may 
depend on what scale and what object we consider. 
Here we adopt the standpoint that the RGEs of $v_a$ 
are those of $\phi_a^0$ \cite{arason} \cite{cvetic} 
\cite{das}. That is, in our case,
\begin{eqnarray}
16\pi^2{dv_1 \over dt}&=&-3{\mathrm{tr}}
\left(Y^{(10)}_uY^{(10)\dagger}_u\right)v_1
-3{\mathrm{tr}}\left(Y^{(10)}_uY^{(126)\dagger}_u\right)v_2 \nonumber\\
&+&\left({9 \over 4}g_2^2 + {3 \over 4}g_Y^2\right)v_1,\\
16\pi^2{dv_2 \over dt}&=&-3{\mathrm{tr}}
\left(Y^{(126)}_uY^{(126)\dagger}_u\right)v_2
-3{\mathrm{tr}}\left(Y^{(126)}_uY^{(10)\dagger}_u\right)v_1 \nonumber\\
&+&\left({9 \over 4}g_2^2 + {3 \over 4}g_Y^2\right)v_2,\\
16\pi^2{dv_3 \over dt}&=&-{\mathrm{tr}}
\left(3Y^{(10)\dagger}_dY^{(10)}_d
+Y^{(10)\dagger}_eY^{(10)}_e\right)v_3 \nonumber\\
&-&{\mathrm{tr}}\left(3Y^{(10)\dagger}_dY^{(126)}_d
+Y^{(10)\dagger}_eY^{(126)}_e\right)v_4 \nonumber\\
&+&\left({9 \over 4}g_2^2 + {3 \over 4}g_Y^2\right)v_3,\\
16\pi^2{dv_4 \over dt}&=&-{\mathrm{tr}}
\left(3Y^{(126)\dagger}_dY^{(126)}_d
+Y^{(126)\dagger}_eY^{(126)}_e\right)v_4 \nonumber\\
&-&{\mathrm{tr}}\left(3Y^{(126)\dagger}_dY^{(10)}_d
+Y^{(126)\dagger}_eY^{(10)}_e\right)v_3 \nonumber\\
&+&\left({9 \over 4}g_2^2 + {3 \over 4}g_Y^2\right)v_4.
\label{v4}
\end{eqnarray}
Here we have replaced the RGEs of $\phi_a$ with $v_a$. 
It goes from Eq.(\ref{mass}) and Eqs.(\ref{yu10})-(\ref{v4}) that
\begin{eqnarray}
16\pi^2{dM_u \over dt}&=&-\frac{3}{\sqrt{2}}\,
{\mathrm{tr}}\left(Y^{(10)}_uY^{(126)\dagger}_u
-Y^{(126)}_uY^{(10)\dagger}_u\right)
\left( v_2 Y^{(10)}_u-v_1 Y^{(126)}_u \right)\nonumber\\
&-&\left(8g_3^2+{2 \over 3}g_Y^2\right)M_u \nonumber\\
&+&{1 \over 2}\left(Y^{(10)}_uY^{(10)\dagger}_u
+Y^{(126)}_uY^{(126)\dagger}_u \right. \nonumber\\
&+&\left.Y^{(10)}_dY^{(10)\dagger}_d
+Y^{(126)}_dY^{(126)\dagger}_d\right)M_u \nonumber\\
&+&M_u\left(Y^{(10)\dagger}_uY^{(10)}_u
+Y^{(126)\dagger}_uY^{(126)}_u\right),\\
16\pi^2{dM_d \over dt}&=&-\left(8g_3^2-{1 \over 3}g_Y^2\right)
M_d \nonumber\\
&+&{1 \over 2}\left(Y^{(10)}_uY^{(10)\dagger}_u
+Y^{(126)}_uY^{(126)\dagger}_u \right. \nonumber\\
&+&\left.Y^{(10)}_dY^{(10)\dagger}_d
+Y^{(126)}_dY^{(126)\dagger}_d\right)M_d \nonumber\\
&+&M_d\left(Y^{(10)\dagger}_dY^{10}_d
+Y^{(126)\dagger}_dY^{126}_d\right),\\
16\pi^2{dM_e \over dt}&=&-3g_Y^2M_e \nonumber\\
&+&{1 \over 2}\left(Y^{(10)}_eY^{(10)\dagger}_e
+Y^{(126)}_eY^{(126)\dagger}_e \right)M_e \nonumber\\
&+&M_e\left(Y^{(10)\dagger}_eY^{(10)}_e
+Y^{(126)\dagger}_eY^{(126)}_e\right).
\end{eqnarray}
It should be remarked that RGEs destroy the 
transpose-invariance of mass matrix possesed at GUT. 
At the region $\Lambda_I \gg \mu$, we must treat the 
RGEs of $\kappa$ of Eq.(\ref{fourpoint}), giving the following forms;
\begin{eqnarray}
16\pi^2\frac{d\kappa^{(1,1)}}{dt}&=&6{\mathrm{tr}}
\left(Y^{(10)}_uY^{(10)\dagger}_u\right)\kappa^{(1,1)} \nonumber\\
&+&3{\mathrm{tr}}\left(Y^{(126)}_uY^{(10)\dagger}_u\right)
\left(\kappa^{(1.2)}+\kappa^{(2.1)}\right) \nonumber\\
&-&3g_2^2\kappa^{(1,1)} \nonumber\\
&+&\frac{1}{6}\left(\lambda_{1111}\kappa^{(1,1)}
+\lambda_{1112}\kappa^{(1,2)}
+\lambda_{1212}\kappa^{(2,2)}\right) \nonumber\\
&+&\frac{1}{2}\left\{\left(Y^{(10)}_eY^{(10)\dagger}_e
+Y^{(126)}_eY^{(126)\dagger}_e\right)\kappa^{(1,1)}\right. \nonumber\\
&+&\left.\kappa^{(1,1)}\left(Y^{(10)}_eY^{(10)\dagger}_e
+Y^{(126)}_eY^{(126)\dagger}_e\right)^T\right\},\\
16\pi^2\frac{d\kappa^{(2,2)}}{dt}&=&6{\mathrm{tr}}
\left(Y^{(126)}_uY^{(126)\dagger}_u\right)\kappa^{(2,2)} \nonumber\\
&+&3{\mathrm{tr}}\left(Y^{(10)}_uY^{(126)\dagger}_u\right)
\left(\kappa^{(1.2)}+\kappa^{(2,1)}\right) \nonumber\\
&-&3g_2^2\kappa^{(2,2)} \nonumber\\
&+&\frac{1}{6}\left(\lambda_{2222}\kappa^{(2,2)}
+\lambda_{2121}\kappa^{(1,1)}\right) \nonumber\\
&+&\frac{1}{2}\left\{\left(Y^{(10)}_eY^{(10)\dagger}_e
+Y^{(126)}_eY^{(126)\dagger}_e\right)\kappa^{(2,2)}\right. \nonumber\\
&+&\left.\kappa^{(2,2)}\left(Y^{(10)}_eY^{(10)\dagger}_e
+Y^{(126)}_eY^{(126)\dagger}_e\right)^T\right\},\\
16\pi^2\frac{d\kappa^{(1,2)}}{dt}&=&3{\mathrm{tr}}
\left(Y^{(10)}_uY^{(10)\dagger}_u
+Y^{(126)}_uY^{(126)\dagger}_u\right)\kappa^{(1,2)} \nonumber\\
&+&3{\mathrm{tr}}\left(Y^{(10)}_uY^{(126)\dagger}_u\right)
\kappa^{(1.1)} \nonumber\\
&+&3{\mathrm{tr}}\left(Y^{(126)}_uY^{(10)\dagger}_u\right)
\kappa^{(2.2)} \nonumber\\
&-&g_2^2\left(2\kappa^{(1,2)}+\kappa^{(2,1)}\right) \nonumber\\
&+&\frac{1}{6}\left(\lambda_{1122}\kappa^{(1,2)}
+\lambda_{1221}\kappa^{(2,1)}\right) \nonumber\\
&+&\frac{1}{2}\left\{\left(Y^{(10)}_eY^{(10)\dagger}_e
+Y^{(126)}_eY^{(126)\dagger}_e\right)\kappa^{(1,2)}\right. \nonumber\\
&+&\left.\kappa^{(1,2)}\left(Y^{(10)}_eY^{(10)\dagger}_e
+Y^{(126)}_eY^{(126)\dagger}_e\right)^T\right\},\\
16\pi^2\frac{d\kappa^{(2,1)}}{dt}&=&3{\mathrm{tr}}
\left(Y^{(10)}_uY^{(10)\dagger}_u
+Y^{(126)}_uY^{(126)\dagger}_u\right)\kappa^{(2,1)} \nonumber\\
&+&3{\mathrm{tr}}\left(Y^{(10)}_uY^{(126)\dagger}_u\right)
\kappa^{(1.1)} \nonumber\\
&+&3{\mathrm{tr}}\left(Y^{(126)}_uY^{(10)\dagger}_u\right)
\kappa^{(2.2)} \nonumber\\
&-&g_2^2\left(\kappa^{(1,2)}+2\kappa^{(2,1)}\right) \nonumber\\
&+&\frac{1}{6}\left(\lambda_{2211}\kappa^{(2,1)}
+\lambda_{2112}\kappa^{(1,2)}\right) \nonumber\\
&+&\frac{1}{2}\left\{\left(Y^{(10)}_eY^{(10)\dagger}_e
+Y^{(126)}_eY^{(126)\dagger}_e\right)\kappa^{(2,1)}\right. \nonumber\\
&+&\left.\kappa^{(2,1)}\left(Y^{(10)}_eY^{(10)\dagger}_e
+Y^{(126)}_eY^{(126)\dagger}_e\right)^T\right\},
\end{eqnarray}
The different factor 2 of the coefficients of $g_2^2$ 
in $\kappa^{(1,2)}$ and $\kappa^{(2,1)}$ comes from the 
different contribution of W-boson and Z-boson loop 
correction on $\kappa^{(1,2)}$ and $\kappa^{(2,1)}$. 
The self coupling contributions come from the diagram Fig.1.\\
Substituting the above equations into Eq.(18), we obtain the 
RGE of the light neutrino mass matrix:
\begin{eqnarray}
16\pi^2\frac{dM_\nu}{dt}&=&\frac{1}{2}
\left\{3(g_2^2+g_Y^2)M_\nu \right.\nonumber\\
&+&\frac{1}{6}\sum_{a,b,c,d=1,2}\lambda_{abcd}v_a^{*}v_c^{*}\kappa^{(b,d)}
\nonumber\\
&+&\left(Y^{(10)}_eY^{(10)\dagger}_e
+Y^{(126)}_eY^{(126)\dagger}_e \right)M_\nu \nonumber\\
&+&\left.M_\nu \left(Y^{(10)}_eY^{(10)\dagger}_e
+Y^{(126)}_eY^{(126)\dagger}_e \right)^T \right\}.
\label{RGEK}
\end{eqnarray}
Thus we have obtained RGEs of SO(10) GUT with two 
Higgs scalars. These formulas are especially important 
for essentially multiple Higgs models where two Higgs 
doblets in up or down type quark-lepton mass matrices 
can not suffer doublet-doublet splittings. 
 Even in a model where the doublet-doublet splitting occurs, 
our results are useful for estimating the threhold effects
 due to the heavy pairs of Higgs doublets.  
We calculated RGE's explicitly in non-SUSY SO(10).  
Those of SUSY SO(10) are obtained analogously 
by considering the contributions of the superpartners of 
non-SUSY contents except for the scalar self couplings.

\section*{Acknowledgments}
We thank N.Okada, Y.Koide, and K.Matsuda for very 
useful arguments. T.F expresses his gratitude to 
R.Mohapatra for hospitality at the University of 
Maryland where a large part of this work was accomplished.

%\end{multicols}

\begin{figure}[htbp]
\begin{center}
\includegraphics{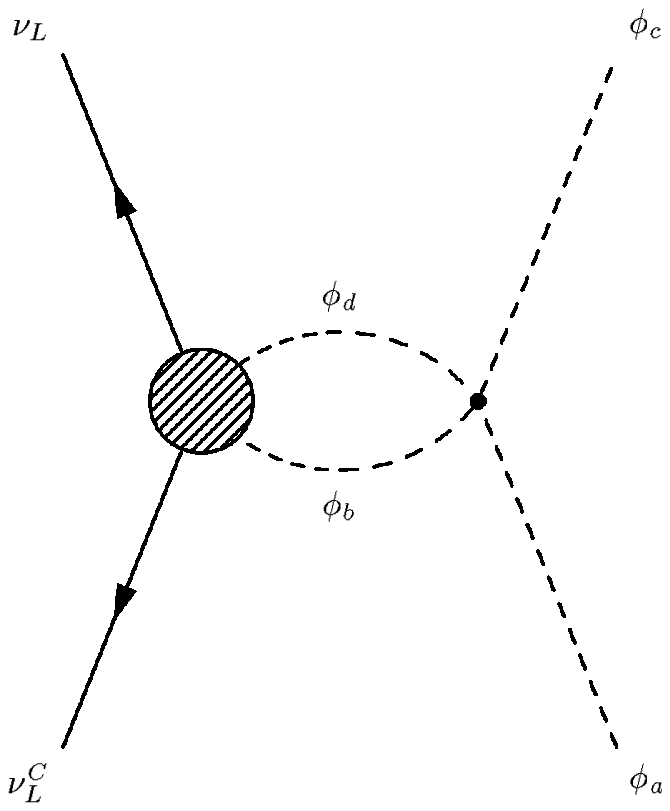}
\end{center}
\caption{The Higgs scalar self couplings contribution 
to the RGE's of $\kappa$.}
\label{fig1}
\end{figure}

\end{document}